\documentstyle[twocolumn,prl,aps,epsfig]{revtex}
\tighten
\newcommand{\beq}{\begin{eqnarray}}
\newcommand{\eeq}{\end{eqnarray}}
\renewcommand{\vec}[1]{{\mathbf{#1}}}
\begin{document}
\draft
\input epsf.sty

\title
{
Fluctuation-Driven First-Order Transition in Pauli-limited 
$d$-wave Superconductors}
\author{Denis Dalidovich and Kun Yang }
\vspace{.05in}

%
\address
{National High Field Magnetic Laboratory and Department of Physics,\\
Florida State University, Tallahassee, Florida 32306, USA}

%
\address{\mbox{ }}
\address{\parbox{14.5cm}{\rm \mbox{ }\mbox{ }
We study the phase transition between
the normal and non-uniform (Fulde-Ferrell-Larkin-Ovchinnikov) 
superconducting state in quasi two-dimensional d-wave superconductors
at finite temperature. We obtain an appropriate Ginzburg-Landau theory 
for this transition, in which the fluctuation spectrum of the order parameter
has a set of minima at non-zero momenta.
The momentum shell renormalization group procedure combined with 
$\varepsilon$ expansion is then applied to analyze the phase 
structure of the theory. 
We find that all fixed points have more than one relevant
directions, indicating the transition is 
of the fluctuation-driven first order type for this universality class.   
}}
\address{\mbox{ }}
\address{\mbox{ }}

\maketitle

It was pointed out forty years ago\cite{ff,lo} that a superconducting 
state with an inhomogeneous order parameter can be stabilized by 
a Zeeman splitting between electrons with opposite spins,
that is comparable to the energy gap. 
This inhomogeneous superconducting state, or 
Fulde-Ferrell-Larkin-Ovchinnikov (FFLO) state, has been the subject of 
mostly theoretical study for many years\cite{review}.
The situation has changed recently
as experimental results suggestive of the FFLO state
start to 
emerge\cite{gloos,modler,tachiki,geg,singleton,radovan,bianchi,martin}.
We would like to mention that
recent experimental results on the heavy fermion compound
CeCoIn$_5$, a quasi two-dimensional (2D) d-wave superconductor,
are particularly compelling\cite{radovan,bianchi,martin}.
It is worth pointing out 
that the FFLO state may also be realized in high
density quark matter, and is thus of interest to the
particle physics community\cite{review}.

Given the long history of the subject matter, it is perhaps somewhat 
surprising that thus far
most of the theoretical studies of the FFLO state are of the
mean-field type. On the other hand one expects that quantum and thermal 
fluctuation effects should be much more significant in FFLO superconductors 
than in ordinary BCS superconductors, as the FFLO phase breaks the 
translational 
symmetry in addition to the gauge symmetry. In a recent work\cite{yang01},
one of us studied the FFLO phase in quasi 1D superconductors, 
and used bosonization to treat intra-chain quantum fluctuations exactly;
one of the conclusions was that the transition from the FFLO phase to the BCS
phase is a continuous transition of the commensurate-incommensurate type, 
in contrast to the first order transition commonly asserted in the literature.
The effect of thermal fluctuations of the superconducting order parameter 
was discussed by Shimahara\cite{shima98}; he argues that the enhanced
fluctuation effects destabilize certain types of mean-field FFLO states.
Anisotropy in pairing 
or electron dispersion may suppress these fluctuation effects 
however\cite{shima98}.

It is perhaps not quite well recognized yet that the FFLO superconductors are 
realizations of the Brazovskii model\cite{brazovskii}, which describes a large
class of statistical mechanical systems in which the fluctuation spectra of
the order parameter have their minima {\em away} from the origin in momentum
space; this is 
precisely the case for the FFLO state, which prefers the superconducting 
order parameter to carry {\em finite} momenta. In its original form,
the Brazovskii model assumes that the fluctuation spectrum has a continuous
set of degenerate minima; it was shown that the fluctuation effects are so
strong that they render the transition between the ordered and disordered
phases a fluctuation-driven first order transition, even if the mean-field
theory suggests a second-order transition. Interestingly, the transition 
between the normal and (possibly) FFLO phase in CeCoIn$_5$ was indeed found
to be first-order\cite{radovan,bianchi}. There are two possible origins for
the first order nature of the transition. Firstly, it is known that near the
tricritical point where the normal, FFLO and BCS superconducting phases 
merge within mean field theory (at $T\approx 0.56 T_c$), the effective 
Ginzburg-Landau free energy has a contribution from quartic terms that
is in certain cases {\em negative}\cite{buzdin,agterb}, 
in which case the mean-field
theory itself predicts a first-order transition. In CeCoIn$_5$ however, the
FFLO phase was observed only at temperatures much lower than the 
tricritical point. It has been pointed out recently\cite{matsuo}
that in the low-temperature regime the quartic term makes {\em positive} 
contributions to the free energy; thus a second-order transition would be 
expected at mean-field level. If this is the case then the origin of the 
first-order transition must be due to fluctuation effects  
left out at the mean-field level. It is this second possibility that we 
focus on in this work. 

While the FFLO phase and the original Brazovskii model share the common feature
that the order parameter spectrum has it minima away from the origin,
one cannot apply the Brazovskii results directly to real systems like
CeCoIn$_5$ because in such systems there is always anisotropy in either
pairing potential or electron dispersion, which would generically reduce the
continuous set of degenerate minima to a discrete set. Thus in this work we
take this important effect into account, and use a model that
contains the anisotropy that is appropriate for a d-wave superconductor with
four-fold symmetry. We perform a renormalization group (RG)
analysis (combined with 
an appropriate $\varepsilon$ expansion, see below), and show 
that the transition is generically first order (at least when $\varepsilon$ is
sufficiently small), even when the mean-field
analysis suggests a second-order transition. In the following we first outline
the derivation of an appropriate Ginzburg-Landau theory from a microscopic
pairing model, then perform the RG analysis.
  

We consider a weak-coupling quasi 2D d-wave superconductor,
whose partition function is
$Z=\int D[\Psi ^\dagger,\Psi] \exp\{-S \}$, with $S=S_0 +S_{int}$, where
 \begin{eqnarray}
&& S_{0} = \sum_{\vec k ,\omega_n} \sum_{\sigma=\uparrow, \downarrow}
  \left[ i\omega_n - \xi(\vec k )-\sigma I \right] 
  \Psi^{\dagger}_{\sigma} (\vec{k},\omega_n)
 \Psi_{\sigma}(\vec{k},\omega_n), \nonumber\\
&& S_{int} = -T \sum_{\vec{k},\vec{k}',\vec{q}} 
   \sum_{\nu,\nu',\omega} V_{\vec{k},\vec{k}'} 
   \Psi^{\dagger}_{\uparrow} (\vec{k}+\vec{q},\nu+\omega)
\nonumber \\ 
&& \cdot \Psi^{\dagger}_{\downarrow}(-\vec{k},-\nu) \Psi_{\uparrow}
(-\vec{k}', -\nu') \Psi_{\downarrow} (\vec{k}' +\vec{q}, \nu'+\omega),
\end{eqnarray}
with $\Psi$,$\Psi^{\dagger}$ being Grassman variables. Index 
$\sigma=\uparrow,\downarrow$ enumerates electron spin, 
and $I$ is the Zeeman splitting that stabilizes the FFLO state\cite{note}.
The electron dispersion
$\xi(\vec{k}) = \epsilon(\vec k_{\|}) +J\cos k_zd - \epsilon_F$,
where $\vec k_{\|}$ denotes the momentum parallel to the planes,
$J$ is the strength of hopping between the layers.
Interaction $V_{\vec{k},\vec{k}'}=V f_{\vec k}f_{\vec{k}}'$ with $V>0$,
is assumed for simplicity to be independent on the $z$-component
of momenta with  $f_{\vec k}=\cos 2\theta_{\vec{k}}$ for the d-wave pairing.
We consider here the clean system only, assuming that disorder does not 
affect qualitatively the phenomena under consideration. 

We decouple then the quartic term via the Hubbard-Stratanovich 
transformation by introducing the complex field $\Delta(\vec q, \omega_n)$,
that serves as a superconducting order parameter.
To obtain the Ginzburg-Landau (GL) action in powers of $\Delta$,
we perform subsequently the cumulant expansion integrating out 
the fermions having
the Green function ($\sigma=\pm$ for up and down spins respectively) 
$G^{-1}(\vec{q},\omega_n)= i\omega_n-\xi(\vec{q})-\sigma I$.
Considering here the finite-$T$ transition only, 
we retain the zero-frequency component $\Delta(\vec q, \omega_n=0)$
in all cumulants. 
The resulting functional takes the form 
\begin{eqnarray}\label{func}
&&{\cal F} =\sum_{\vec q} K_2(\vec q) |\Delta(\vec q )|^2 \nonumber\\
&&+\sum\nolimits_{\vec q_1,..,\vec q_4}^{'}
 K_4 (\vec q_1,..,\vec q_4)
\Delta(\vec q_1)\Delta(\vec q_2)\Delta^* (\vec q_3)\Delta^* (\vec q_4), 
\end{eqnarray}
where the prime in the sum over momenta in the quartic 
term indicates that the condition $\vec q_1+\vec q_2-\vec q_3-\vec q_4=0$
is taken into account.
Since the transition occurs into the state that 
is non-uniform in space, the full momentum dependence of $K_2$ and $K_4$ must 
be kept\cite{note2}.
$K_2(\vec q )$ is given by the standard bubble 
diagram with two external legs,
$K_2(\vec q)=1/V-T\sum_{\vec k, \nu_n} 
f_{\vec k}^2
G_{\uparrow}\left(\vec k +\vec q, \nu_n \right)
\times G_{\downarrow}\left(-\vec k, -\nu_n \right)$.
In this formula, one performs then straightforwardly the summation over 
frequency and integration over momentum within the shell around the 
Fermi surface $|k-k_F|\le \omega_D/v_F (\vec k_F)$.
The form of the Fermi surface is assumed to have the same 4-fold
d-wave symmetry
in the $k_{\|}$ plane. 
Distinguishing also the components of the Fermi-momentum
parallel and perpendicular to the planes, 
$\vec k_F=(\vec k^{(F)}_{\|}, k^{(F)}_{z})$, 
$\vec v_F=(\vec v_{\|}, v_z)$, we find as a result that 
for $q\ll k_F$\cite{kyang} 
\beq\label{K2}
&&K_2(\vec q)=\frac{1}{V}-
\frac{1}{(2\pi)^3}\int_{-\pi/d}^{\pi/d} 
\frac{k^{(F)}_{\|}dk_z}{v_F (\vec k_F)} \int  d\theta
\cos^2 2\theta \nonumber\\
&&\int_{0}^{\omega_D} \frac{d\epsilon}{2\epsilon}
\left[ \tanh\left( \frac{\epsilon+z_{\vec q}}{2T}\right)+
\tanh\left( \frac{\epsilon-z_{\vec q}}{2T}\right) \right].
\eeq 
In the equation above, $\hbar=1$, 
\beq
z_{\vec q}= \frac12\left[ v_{\|} q_{\|}\cos(\theta_{\vec v}-\theta_{\vec q})
 - Jd q_z \cdot \sin k_z d \right] +I,
\eeq
with $\theta$, $\theta_{\vec v}$ and $\theta_{\vec q}$ being the 
in-plane angles
of $\vec k$, $\vec v_F$ and the pairing momentum $\vec q$ respectively. 
$\vec k_F$, $\vec v_F$ as well as $\theta_{\vec v}$,
are themselves functions of $k_z$ and $\theta$ 
characterizing the Fermi surface.

To determine the absolute value of the pairing momentum $q_0$ and its
direction, it is necessary to find the minima of $K_2(\vec q)$ with respect 
to $q_{\|}$ and $\theta_{\vec q}$, as well as $q_z$.
It is clear that the ordering wave vector must lie in the $(q_x,q_y)$
plane meaning that $q_{0z}=0$. 
However, investigation of $K_2(\vec q)$ regarding the minimum with respect 
to $\theta_{\vec q}$, reveals that Eq. (\ref{K2}) has extrema for 
the two sets of values:
$\theta_i=\frac{\pi}{4}, \frac{3\pi}{4}, \frac{5\pi}{4}, -\frac{\pi}{4}$
and $\theta_i=0, \frac{\pi}{2}, \pi, \frac{3\pi}{2}$
correspond to the nodal and anti-nodal directions in the $(q_x,q_y)$ plane.
Generally speaking, for each of these sets one obtains the different values
of $q_0$ as a result of the solution of equation 
$\partial K_2(\vec q)/\partial q_0$.  
Whether both of these sets minimize $K_2(\vec q)$, or only one of them 
is actually a minimum with the other being the maximum, is determined by 
the specific form of the Fermi surface. It is important that, if both
sets with the corresponding
values of $q_0$ are the minima, the actual
transition will occur into configuration described by the set
leading to the largest critical value $T_c$\cite{kyang,maki}. 
In case of the anti-nodal ordering, it is possible to expand 
$K_2(\vec q)$ for $\vec q$ located in the pockets near
the minima $q_{0x}^{(i)}$,$q_{0y}^{(i)}$, corresponding
to the direction $\theta_i$,
$K_2(\vec q)=r+\alpha_x^{(i)} (q_{x}-q_{0x}^{(i)})^2+
\alpha_y^{(i)} (q_{y}-q_{0y}^{(i)})^2+ \gamma q_z^2$, where 
$\alpha_x^{(i)}$ and $\alpha_y^{(i)}$ mutually interchange for the 
neighboring pockets. For the case of nodal ordering, one can show
from Eq. (\ref{K2}) that $K_2(\vec q)=r+\alpha (q_{x}-q_{0x}^{(i)})^2+
2\beta^{(i)} (q_{x}-q_{0x}^{(i)})(q_{y}-q_{0y}^{(i)})+
\alpha (q_{y}-q_{0y}^{(i)})^2+ \gamma q_z^2$, 
where $\beta^{(i)}$ are opposite in sign for the neighboring pockets.
By the simple rotation of coordinate system in the
$(q_x,q_y)$ plane by $\pi/4$, the latter expansion, however, reduces to that
for the anti-nodal case.

The quartic kernel in Eq. (\ref{func})
is given by the bubble containing four electron Green 
functions and four external legs representing the order parameter field
$\Delta(\vec q)$, 
\beq\label{K4}
&&K_4 = (T/4)\sum_{\nu_n,\vec k} \left\{ f^2_{\vec k}
f^2_{\vec k+\vec q_4-\vec q_2}
G_{\uparrow}(\vec k +\vec q_1, \nu_n)
G_{\downarrow}(-\vec k, -\nu_n)\right. \nonumber\\ 
&&\left.G_{\uparrow}(\vec k +\vec q_4, \nu_n)
G_{\downarrow}(-\vec k+\vec q_2-\vec q_4, -\nu_n)
+\left[ \vec q_4 \rightarrow \vec q_3 \right] \right\},
\eeq
where $[\vec q_4 \rightarrow \vec q_3 ]$ stands for the same expression as 
right before, with only $\vec q_4$ substituted by $\vec q_3$.
The kinematic constraint
$\vec q_1+\vec q_2=\vec q_3+\vec q_4$
is implied in Eq. (\ref{K4}).
It will not be required, however, to know this cumulant  
for all values of momentum variables. 
Since we are considering the renormalization group treatment involving
the wave-vectors located in the pockets near $\vec q_0^{(i)}$, 
only those values in Eq. (\ref{K4})
are of interest, in which $\vec q$'s point right to the centers of the 
pockets and satisfy the aforementioned constraint .  
The following distinct possibilities can be readily enumerated, once 
one denotes by $i$ the number of the pocket in the
$\vec q_{\|}$ plane; $i$ equals to $1,2,3,4$ starting from that 
with the lowest value of angle, with the formal condition $i+4=i$.
\beq\label{inta}
&&K_4( \vec q_0^{(i)},\vec q_0^{(i)},\vec q_0^{(i)},\vec q_0^{(i)})=
u_0/4,i=1,...,4;
\eeq
\beq\label{intb}
&&K_4( \vec q_0^{(i)},\vec q_0^{(i+1)},\vec q_0^{(i)},\vec q_0^{(i+1)})=
u_{\pi/2}/4, i=1,...,4;
\eeq
\beq\label{intc}
&&K_4( \vec q_0^{(i)},\vec q_0^{(i+2)},\vec q_0^{(i)},\vec q_0^{(i+2)})=
u_{\pi}/4,\quad i=1,2;
\eeq
\beq\label{intd}
&&K_4( \vec q_0^{(1)},\vec q_0^{(3)},\vec q_0^{(2)},\vec q_0^{(4)})=
v/4.
\eeq
Looking at the expressions given by Eqs. (\ref{inta})-(\ref{intd}), we 
see that $u_0$ describes the interactions between the modes within
the same pocket, while the other parameters account for the inter-pocket
scattering. In parts with $u_{\pi/2}$ and $u_{\pi}$, the interaction
occurs between the pockets that have the angles between their 
$\vec q_0$'s equal to $\pi/2$ and $\pi$ respectively.  
Without providing the explicit expressions for 
those coefficients, we note that the only point important 
in the general derivation is that all interactions are non-singular
at the critical values $T_c$ and $\vec q_0$.
The interactions can in principle have arbitrary signs 
that may change along the critical line $T_c=T_c(I)$ on the $(T, I)$ plane.
For example, as it was mentioned in the introduction,
interaction $u_0$ is negative close to the tricritical point,
where the normal, uniform and non-uniform superconducting phases meet
\cite{agterb}. At the same time, at lower temperatures $u_0$ seems to be 
positive\cite{review}. If $u_0<0$, the transition is necessarily first 
order already at the mean-field level. 
Hence, we will assume that we consider the transition only in those 
regions on the phase diagram, in which at least $u_0$ is positive. 

To distinguish the modes with the wave vectors belonging to the
different pockets, it is convenient to introduce the shifted momenta 
$\vec k= \vec q-\vec q_0^{(i)}$ and decompose the total field 
$\Delta (\vec q)$ into the parts $\Delta_i (\vec k=\vec q-\vec q_0^{(i)})$.
Each part $\Delta_i (\vec k)$ accounts for the fluctuations having the 
momenta in the vicinity of $\vec q_0^{(i)}$. 
It is clear that under such decomposition,
the kinematic constraint $\vec k_1+\vec k_2=\vec k_3+\vec k_4$ 
for the shifted vectors is preserved. Though $\vec k_j$ in the arising 
quartic terms generally belong to the different
pockets, one can treat them during the formulation of RG equations, 
as if they are located in one and the same pocket 
around the origin. We will use in RG equations below the form of the 
propagator obtained for the anti-nodal ordering assuming
for clarity that $\alpha_x^{(1)}=\alpha_1 \ne \alpha_y^{(1)}=\alpha_2$,  
meaning the spatial anisotropy in the spectrum of excitations.
The issue of spatial anisotropy in RG near quantum critical points was 
addressed in different physical context in Ref.\cite{vojta}, albeit the 
anisotropy there was related to fermionic excitations. 

After the appropriate rescaling of momentum variables and fields, 
the general GL action takes the form:
\beq\label{GL}
&&{\cal F}=\sum_i \sum_{\vec k} \left[ r+ \alpha_x^{(i)} k_x^2
+ \alpha_y^{(i)} k_y^2 + k_z^2 \right]
|\Delta_i (\vec k)|^2 \nonumber\\
&&+\sum\nolimits_{ \{ k_j \} }^{'}
\left\{ (u_{0}/4)\sum_i 
\Delta_i (\vec k_1) \Delta_i (\vec k_2) \Delta_i^* (\vec k_3)
\Delta_i^* (\vec k_4)\right.\nonumber\\
&&\left. +u_{\pi/2}\sum_{[i]} 
\Delta_i (\vec k_1) \Delta_{i+1} (\vec k_2)\Delta_i^* (\vec k_3)
\Delta_{i+1}^* (\vec k_4) \right.\nonumber\\
&&\left. + u_{\pi}\sum_{i=1,2}
\Delta_i (\vec k_1) \Delta_{i+2} (\vec k_2)\Delta_i^* (\vec k_3)
\Delta_{i+2}^* (\vec k_4) \right.\nonumber\\
&&\left.+v \left( \Delta_1 (\vec k_1) \Delta_3 (\vec k_2) 
\Delta_2^* (\vec k_3)\Delta_4^* (\vec k_4)+ {\rm c.c } \right) \right\}.
\eeq 
A few more remarks on the notations in Eq. (\ref{GL}) are in order.
The notation $\sum_{ \{ k_j \} }^{'}$ implies that the summation over
$\vec k_j$ is taken with the restriction
$\vec k_1+\vec k_2=\vec k_3+\vec k_4$.
$\sum_{[i]}$ means that the sum over $i$ is performed
with the condition $i+4=i$. In all terms of the quartic part, 
except that with $u_0/4$, the permutational symmetry between the fields
arising from the obvious permutations of momenta in arguments of 
Eqs. (\ref{intb})-(\ref{intd}), is taken care of explicitly, canceling 
thus the factor of $4$ in denominator. This greatly simplifies the 
subsequent RG loop analysis. 

Couplings $u_0$ and $u_{\pi}$ are in fact the primary parameters, 
whose flow under rescaling determines the character of transition.  
To see this, we calculate the free energy at the mean-field level for  
two possible phases:
1)Fulde-Ferell (FF) phase with 
$\Delta(\vec r)=\Delta_0 e^{i\vec q \vec r}$ and
2)Larkin-Ovchinnikov phase having
$\Delta(\vec r)=\Delta_0 \cos(\vec q \cdot \vec r)$. The values are
${\cal F}_{{\rm FF}}=-|r|^2/u_0$,
${\cal F}_{{\rm LO}}=-2|r|^2/(u_0+2u_{\pi})$.
The LO phase has the lower energy when $u_0 > 2u_{\pi}$, while
the  FF phase is more favorable under the opposite
condition. 
The considerations above necessarily imply that $u_0 >0$, since
only in this case the transition is of the second order at the mean field 
level. In addition, if LO phase is realized, one must ensure that not
only $u_0>0$ but also $u_0+2u_{\pi} >0$. Those requirements will be presumed
fulfilled in the subsequent treatment.
  
Simple tree-level scaling applied to Eq. (\ref{GL}) 
shows that if the effective dimensionality of the problem, $d>d_c=4$, 
the interactions are irrelevant and the transition is of the second order.
To proceed, we will generalize the $z$-component of momentum to 
$k_{\perp}$ having the dimensionality $2-\varepsilon$, and integrate
out the modes in the thin layer
$\Lambda /e^l <k_x,k_y <\Lambda$
around the square shell: $-\Lambda <k_x <\Lambda$, $k_y=\pm\Lambda$;
$-\Lambda <k_y <\Lambda$, $k_x=\pm\Lambda$, with the integrals 
over $k_{\perp}$ taken from $-\infty$ to $\infty$.
The arising one-loop RG equations for interactions are:
\beq\label{u0ren}
du_0/dl=\varepsilon u_0- 2f \left[ (5/4) u_0^2 + 
2u_{\pi/2}^2  +u_{\pi}^2 \right],
\eeq
\beq\label{upiren}
du_{\pi}/dl=\varepsilon u_{\pi} - 2f \left[
u_0 u_{\pi} + u_{\pi}^2  
+u_{\pi/2}^2  + v^2/2 \right],
\eeq
\beq\label{upi/2ren}
\frac{du_{\pi/2}}{dl}=\varepsilon u_{\pi/2}- 2
\left[ f u_{\pi/2} (u_0+u_{\pi}) 
+g(u_{\pi/2}^2 +\frac{v^2}{2}) \right],
\eeq
\beq\label{vren}
dv/dl=\varepsilon v  - 2 v \left[
f u_{\pi} + 2 g u_{\pi/2}\right],
\eeq
with
\beq
f=\frac{4}{(2\pi)^3}\int_{-\infty}^{\infty}
k_{\perp}dk_{\perp} \left[ \int_0^{\Lambda} dk_x
{\cal G}^2(\alpha_1 k_x^2,\alpha_2 \Lambda^2)\right. \nonumber\\
\left. + \int_0^{\Lambda} dk_y
{\cal G}^2(\alpha_1 \Lambda^2,\alpha_2 k_y^2)
\right]\nonumber,
\eeq
\beq
&&g=\frac{4}{(2\pi)^3}\int_{-\infty}^{\infty}
k_{\perp}dk_{\perp} \left[ \int_0^{\Lambda} dk_x
{\cal G} (\alpha_1 k_x^2,\alpha_2 \Lambda^2)\cdot \right. \nonumber\\
&&\left. {\cal G} (\alpha_2 k_x^2,\alpha_1 \Lambda^2)
+ \int_0^{\Lambda} dk_y 
{\cal G} (\alpha_1 \Lambda^2,\alpha_2 k_y^2)
{\cal G} (\alpha_2 \Lambda^2,\alpha_1 k_y^2)\right]\nonumber.
\eeq
In the equations above, ${\cal G}(x,y)=1/(k_{\perp}^2+x+y)$, and we set
$r(l)\sim O(\varepsilon)$ to zero in ${\cal G}(x,y)$ in the one-loop
approximation. Looking for the fixed points, we absorb $f$ 
by rescaling the interactions, generating thus the anisotropy parameter
$a=g/f= (2\sqrt{\eta}/(\eta-1))\arcsin (\eta-1)/(\eta+1)$, where
$\eta={\rm max} \{\alpha_1/\alpha_2, \alpha_2/\alpha_1 \}>1$. This parameter
is cutoff-independent and satisfies $0< a \le 1$.
As it can be seen from Eq. (\ref{vren}), it is 
reasonable to search separately the fixed points that have 
$v^*= 0$ and $v^*\ne 0$.
We have found that setting the condition $v^* \ne 0$,
leads to the absence of fixed points in the space of real variables
for $0< a \le 1$.
Concerning the fixed points having $v^*=0$, it can be shown that
apart from the completely unstable Gaussian fixed point
one has four more points:
$u_{\pi/2}^*=u_{\pi}^*=0, u_0^* =2\varepsilon/5$;
$u_{\pi/2}^*=0, u_{\pi}^* =u_0^*/2 =\varepsilon/6$;
plus two more fixed points that are some cumbersome functions of $a$ 
not to be presented here.
For $a=1$, one easily finds the latter to be
$u_{\pi/2}^* =u_{\pi}^* =u_0^*/2 =\varepsilon/8$,
$u_{\pi/2}^* =u_{\pi}^* =u_0^*/6=\varepsilon/16$;
while for $a\rightarrow 0$ they both collapse onto the point
$u_{\pi/2}^*=0, u_{\pi}^*=u_0^*/2=\varepsilon/6$.
All of the found this way fixed points are unstable,
since, as follows from Eq. (\ref{vren}), there will be at least one 
direction with the eigenvalue
$\lambda_v=\varepsilon-4u_{\pi/2}^*-2u_{\pi}^*$,  
{\it positive} at all the fixed points in the whole range $0<a \le 1$. 
We thus find no stable fixed points at 
the one-loop level, meaning that one needs to tune at least two parameters
($r$ and $v$) to reach the fixed points. This implies that the transition will 
be generically of the 
first order, even if the mean-field theory suggests a second-order transition. 
The situation here is  similar to that near transitions described by the 
effective Hamiltonians 
of anisotropic systems\cite{mukamel}.

In summary, we have obtained an effective Ginsburg-Landau theory for
the transition from normal to the FFLO
state in quasi 2D d-wave superconductors at all non-zero 
temperatures. RG analysis of the theory
indicates that the transition is generically first order, even when the 
mean-field theory suggest a continuous transition. This fluctuation-driven 
first order transition is due to the enhanced fluctuations of the FFLO state, 
associated with additional broken symmetries. Our result is consistent with
the first order character of the transition observed in CeCoIn$_5$.

Part of this work was performed while one of the authors (KY) was 
visiting Kavli Institute of Theoretical Physics; he thanks Subir Sachdev 
for helpful discussions.
This work was supported by the State of Florida (DD), and the NSF
Grants No. DMR-0225698 (KY) and PHY-9907949 (at KITP).


\begin{thebibliography}{99}
\bibitem{note} In real systems $I$ can be due to either (i) an external 
magnetic 
field or (ii) an internal field if the system is ferromagnetic. In the case (i)
there is also an orbital effect; however for the field that is parallel to the 
planes of a  quasi 2D system the orbital effect is weak. In the following we 
neglect the orbital effect, so strictly speaking our results apply to case (ii)
but should also be relevant to case (i) as well.
\bibitem{note2}
The approach used here to obtain the effective GL action is 
different from previous work\onlinecite{buzdin,agterb}, 
which used a gradient expansion (assuming
small order parameter momentum) that is justified near the tricritical point 
only; here we 
have kept the full momentum dependence of the kernels, which is essential to 
correctly identify the optimal momentum of the order parameter at low $T$.
\end{thebibliography}
\end{document}